\documentclass[sigconf]{acmart}
\usepackage{amsmath}
\usepackage{amssymb}
\usepackage{mathrsfs}
\usepackage{algorithm,algpseudocode}
\usepackage{color,xcolor}
\usepackage{algpseudocode}
\usepackage{amsmath}
\usepackage{graphics}
\usepackage{epsfig}
\usepackage{balance}
\usepackage{threeparttable}
\usepackage{enumitem}
\usepackage{bm}
\usepackage{verbatim}
\usepackage{soul}
\usepackage{multirow}
\usepackage{makecell}
\usepackage{xspace}
\usepackage{appendix}
\usepackage{booktabs}
\usepackage{multirow}

\newtheorem{myDef}{Definition}

\newcommand{\eg}{\emph{e.g.,}\xspace}
\newcommand{\ie}{\emph{i.e.,}\xspace}
\newcommand{\wrt}{\emph{w.r.t.}\xspace}
\newcommand{\model}{ATBRG}

\AtBeginDocument{%
  \providecommand\BibTeX{{%
    \normalfont B\kern-0.5em{\scshape i\kern-0.25em b}\kern-0.8em\TeX}}}

\copyrightyear{2020}
\acmYear{2020}
\setcopyright{acmcopyright}\acmConference[SIGIR '20]{Proceedings of the 43rd International ACM SIGIR Conference on Research and Development in Information Retrieval}{July 25--30, 2020}{Virtual Event, China}
\acmBooktitle{Proceedings of the 43rd International ACM SIGIR Conference on Research and Development in Information Retrieval (SIGIR '20), July 25--30, 2020, Virtual Event, China}
\acmPrice{15.00}
\acmDOI{10.1145/3397271.3401428}
\acmISBN{978-1-4503-8016-4/20/07}

\begin{document}

\title{\model: Adaptive Target-Behavior Relational Graph Network for Effective Recommendation}

\author{Yufei Feng$^{1\ast}$, Binbin Hu$^{2\ast}$, Fuyu Lv$^{1}$, Qingwen Liu$^{1}$, Zhiqiang Zhang$^{2}$, Wenwu Ou$^{1}$}
\thanks{$^{\ast}$ indicates both authors contribute equally to this work.}
\affiliation{
 \institution{$^{1}$Alibaba Group, Hangzhou, China\\$^{2}$Ant Financial Services Group, Hangzhou, China}
}
\email{fyf649435349@gmail.com, bin.hbb@antfin.com, {fuyu.lfy, xiangsheng.lqw}@alibaba-inc.com, lingyao.zzq@antfin.com, santong.oww@taobao.com}

\begin{abstract}
Recommender system (RS) devotes to predicting user preference to a given item and has been widely deployed in most web-scale applications. Recently, knowledge graph (KG) attracts much attention in RS due to its abundant connective information. Existing methods either explore independent meta-paths for user-item pairs over KG, or employ graph neural network (GNN) on whole KG to produce representations for users and items separately. Despite effectiveness, the former type of methods fails to fully capture structural information implied in KG, while the latter ignores the mutual effect between target user and item during the embedding propagation. In this work, we propose a new framework named \textbf{A}daptive \textbf{T}arget-\textbf{B}ehavior \textbf{R}elational \textbf{G}raph network ({\model} for short) to effectively capture structural relations of target user-item pairs over KG. Specifically, to associate the given target item with user behaviors over KG, we propose the graph connect and graph prune techniques to construct adaptive target-behavior relational graph. To fully distill structural information from the sub-graph connected by rich relations in an end-to-end fashion, we elaborate on the model design of {\model}, equipped with relation-aware extractor layer and representation activation layer. 
We perform extensive experiments on both industrial and benchmark datasets.
Empirical results show that {\model} consistently and significantly outperforms state-of-the-art methods. Moreover, {\model} has also achieved a performance improvement of 5.1\% on CTR metric after successful deployment in one popular recommendation scenario of Taobao APP. 

\end{abstract}



\maketitle

\section{Introduction}
\begin{figure*}
    \centering
    \includegraphics[scale=0.3]{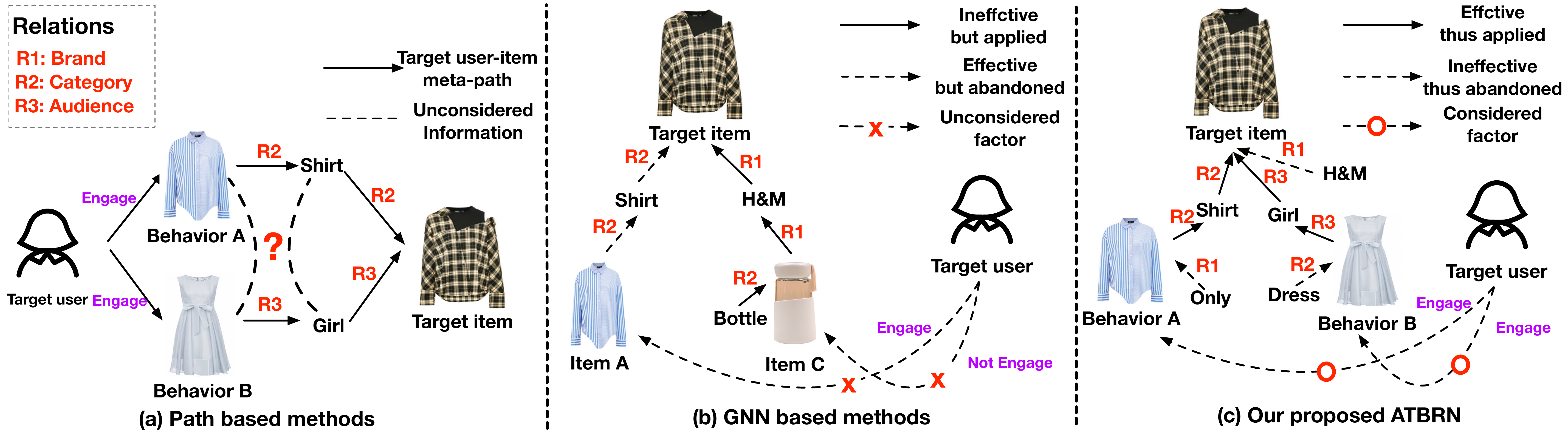}
    \caption{The comparison of our proposed framework  {\model} with previous models. (a) \& (b) indicate the limitations of path based and GNN based methods, respectively, while (c) shows the superiority of {\model} .}
    \label{fig:intro_pic}
\end{figure*}
In the era of information overload, recommender system (RS), which aims to match diverse user interests with tremendous resource items, are widely deployed in various online services, including e-commerce~\citeN{Sarwar:ItemCF, Zhou:DIN, xu2019relation}, social media~\citeN{Paul:YoutubeNet, zhou2019real} and news~\citeN{das2007google, wang2018dkn}. Traditional recommendation methods, \eg matrix factorization ~\cite{koren2009matrix}, mainly learn an effective preference prediction function using historical user-item interaction records. Despite effectiveness, these methods suffer from cold-start problem due to data sparsity. With the rapid development of web services, some approaches~\cite{Hu:McRec,huang2018improving} are proposed to incorporate various auxiliary data for improving recommendation performance. 

Recently, knowledge graph (KG), which is flexible to model comprehensive auxiliary data, has attracted increasing attention in RS~\cite{Wang:KPRN,Wang:KGCN,Wang:KGAT,wang2018ripplenet,cao2019unifying,wang2019explainable}. 
Generally, KG stores external heterogeneous knowledge in the ternary form $\langle head \ entity, relation, tail \ entity \rangle$, corresponding to attribute (\eg $\langle Blouse, Category, Shirt \rangle$ or relationship (\eg $\langle Shirt, Audience, Girl \rangle$) of entities. 
Due to its abundant information, current recommender systems mainly aim to incorporate KG to enrich representations of users and items and promote the interpretability of recommendations.

Though with great improvements, it remains challenging to effectively integrate such heterogeneous information for recommendation. Roughly speaking, state-of-the-art KG based recommendation methods mainly fall into two groups, path based and graph neural network (GNN) based methods. Path based methods~\cite{Wang:KPRN} infer user preference by exploring multiple  meta-paths for target user-item pairs over KG, which typically requires domain knowledge. More importantly, this type of methods ignores rich structural information implied in KG, and thus cannot sufficiently characterize the underlying relationships between given target user and item. As illustrated in Fig.~\ref{fig:intro_pic}, these methods essentially overlook the strong relationships between \emph{Blouse} and \emph{Dress}, since each extracted path is modeled independently.

Inspired by the recently emerging graph neural networks, several GNN based methods~\citeN{Wang:KGCN,Wang:KGAT} have been proposed and provide strong performance by explicitly modeling high-order connectivities in KG. Nevertheless, these methods still suffer from three limitations: (\textbf{L1}) These methods mainly apply GNN to enrich the representation of target user and item separately by aggregating their own original neighbors in the KG, and thus fail to capture their mutual influence during the procedure of information aggregation. As shown in Fig.~\ref{fig:intro_pic} (b), current GNN based methods tend to produce representation for target item through aggregating its neighbors without considering target user's interests (history behaviors). Subsequently, some unnecessary information (\ie \emph{Cup}) is involved in target item's refined embedding, which may harm recommendation performance; (\textbf{L2}) The KG in the real-world industrial scenario is extremely large-scale, where one entity can be linked with up to millions of items. Existing works mainly employ the random sampling on the neighbors beforehand, which may lose latent critical information for the specific target user and item. As shown in Fig.~\ref{fig:intro_pic} (b), some neighbors (\ie \emph{Shirt}) are abandoned by the random sampling strategy, while they are usually informative during aggregation since the target user has engaged with them;
(\textbf{L3}) Most of these methods neglect rich relations among user behaviors over KG, while some works \citeN{Feng:DSIN, zhou2019deep} have  demonstrated that capturing the relations among user behaviors is also beneficial for expressing user preference.

To address above limitations, 
we aim to distill the original over-informative KG into recommendation in a more effective way, which is expected to satisfy the following key properties:
(1) \textbf{Target-behavior}: we hang on the novel insight that an effective KG base recommendation should produce semantic sub-graph to adapt for each target user-item pair, with the aim of capturing the underlying mutual effect characterized by KG (\textbf{L1});
(2) \textbf{Adaptive}: distinct from random sampling on the whole KG, our idea is to follow the adaptive principle for the sub-graph construction, which adaptively preserves useful information connecting user behaviors and target item over the KG, driving our model to provide more effective recommendation (\textbf{L2});
(3) \textbf{Relational}: the model architecture should be designed to relation-aware in order to consider the rich relations among user behaviors and target item over KG (\textbf{L3}). 
For convenience, given a target user-item pair, we call the relational graphical structure bridging user behaviors (\ie historical click records) with the target item as \emph{adaptive target-behavior relational graph} (shown in Fig.~\ref{fig:intro_pic}(c)). Propagating user preference on such a relational structure potentially takes full advantage of the mutual effect for target user-item pair, as well as comprehensively captures the structural relations derived from KG.

In this paper, by integrating above main ideas together, we propose a new framework named \textbf{A}daptive \textbf{T}arget-\textbf{B}ehavior  \textbf{R}elational \textbf{G}raph \textbf{N}etwork ({\model}), which is comprised of two main parts: 
(1) Graph construction part. To extract the effective relational sub-graph of for target the user-item pair over the KG adaptively, we propose the \textit{graph connect} and \textit{graph prune} techniques. Firstly, we explore multiple layer neighbors over KG for target item and each item in user behavior, respectively. Among these entity sets, we \textit{connect} entities which appear in multiple entity sets and \textit{prune} entities belonging to only one entity set.
Subsequently, we construct the adaptive target-behavior relational graph, which characterizes the structural relations among user behaviors and target item over the KG. (2) Model part. Considering structural relations derived from KG, we technically design the relation-aware extractor layer, which employs relation-aware attention mechanism to aggregate structural knowledge over the relational graph for each user behavior and target item. Afterwards, we introduce the representation activation layer to activate the relative relational representations of user behavior \wrt that of target item. 

The main contributions of our work are summarized as follows:
\begin{itemize}
    \item To effectively characterize structural relations between the given target user and item, we propose to extract an adaptive target-behavior relational graph, where the graph connect and graph prune strategies are developed to adaptively build relations between user behaviors and target item over KG.
    \item We propose a novel framework {\model}, a well-designed graph neural network based architecture to learn relational representations of user behaviors and target item over the extracted sub-graph. Moreover, we equip it with relation-aware extractor layer and representation activation layer for emphasizing rich relations for interaction in KG.
    \item We perform a series of experiments on a benchmark dataset from Yelp and an industrial dataset from Taobao App. Experimental results demonstrate that {\model} consistently and significantly outperforms various state-of-the-art methods. Moreover, {\model} has been successfully deployed in one popular recommendation scenario of Taobao APP and gained a performance improvement of 5.1\% on CTR metric.
\end{itemize}
\section{Related Work}
In this section, we review the most related studies in behavior based and knowledge aware recommendation.

\subsection{Behavior based recommendation} \label{URP}
In the early stage of recommendation, researchers focus on recommending a suitable list of items based on historical user-item interaction records. In particular, a series of matrix factorization based methods~\cite{koren2009matrix} have been proposed to infer user preference towards items through learning latent representations of users and items. Due to the ability of modeling complex interaction between users and items, deep neural network based methods (\eg YoutubeNet~\cite{Paul:YoutubeNet}, DeepFM~\cite{Guo:DeepFM}) are widely adopted in industrial recommender systems, and reveal the remarkable strength of incorporating various context information (\eg user profile and item attributes).  

In the online e-commerce systems, we are particularly interested in user's historical behaviors, which implies rich information for inferring user preference. Hence, how to effectively characterize the relationships between user behaviors and target item remains a continuous research topic.
DIN~\cite{Zhou:DIN} adaptively learns the representation of user interests from historical behaviors w.r.t. the target item by the attention mechanism. 
Inspired by DIN, the majority of following up works inherit this kind of paradigm. 
GIN~\cite{Li:GIN} mines user intention based on co-occurrence commodity graph in the end-to-end fashion. 
ATRANK~\cite{Zhou:ATRank} proposes an attention-based behavior modeling framework to model users' heterogeneous behaviors. 
DIEN~\cite{zhou2019deep} and SDM \cite{lv2019sdm} devote to capturing users' temporal interests and modeling their sequential relations.
DSIN~\cite{Feng:DSIN} focuses on capturing the relationships of  users' inter-session and intra-session behaviors.
MIMN~\cite{Pi:MIMN} and HPMN~\cite{ren2019lifelong} apply the neural turing machine to model users' lifelong sequential behaviors. 
Besides these improvements, knowledge graph, consisting of various semantics and relations, emerges as an assistant to describe relationships between user behaviors and target item.

\subsection{Knowledge Aware Recommendation} \label{KAR}
As a newly emerging direction, knowledge graph is widely integrated into recommender systems for enriching relationships among user behaviors and
items. A research line utilizes KG aware embeddings (\eg structural embeddings~\cite{zhang2016collaborative} and semantics embeddings~\cite{wang2018dkn}) to enhance the quality of item representations. These methods conduct mutli-task learning within two tasks of recommendation and KG completion and share the embeddings, and thus can hardly take full advantage of high-order information over KG. On the contrary, several efforts~\cite{Hu:McRec,Wang:KPRN} have been made to explore different semantic path (meta-path) connecting target users and items over KG, and then learn prediction function through multiple path modeling. More recently, some works~\cite{xian2019reinforcement} propose to exploit reinforcement learning to explore useful path for recommendation. Despite effectiveness, the path based method ignores rich structural information implied in KG since each extracted path is modeled independently.

Recently, graph neural network has shown its potential in learning accurate node embeddings with the high-order graph topology. Taking advantages of
information propagation, RippleNet~\cite{wang2018ripplenet} propagates users' potential preferences and explores their hierarchical interests over KG, while KGCN-LS~\cite{Wang:KGCN} and KGAT~\cite{Wang:KGAT} perform embedding propagation by stacking multiple KG aware GNN layers. Although GNN based methods have achieved performance improvement to some extent, they do not take mutual influence between target user behaviors and item into consideration in the procedure of information aggregation.
Moreover, the exponential neighborhood expansion over graph extremely increases the complexity of the system.

\section{Preliminary}
In a recommendation scenario (\eg e-commerce and news), we typically have a series of historical interaction records (\eg purchases and clicks) between users and items. Let $\mathcal{U}$ denote a set of users and $\mathcal{I}$ denote
a set of items, we denote interaction records as 
$\mathcal{H} = \{(u, i, \mathcal{B}_{ui}, y_{ui} | u \in \mathcal{U}, i \in \mathcal{I}\}$. Here, $\mathcal{B}_{ui} \subset \mathcal{I}$ represents historical behaviors (\ie item list) for user $u$ when recommending item $i$ and $y_{ui} \in \{0, 1\}$ is the implicit feedback of user $u$ \wrt item i, where $y_{ui} = 1$ when $\langle u,i \rangle$ interaction is observed, and $y_{ui}$ = 0 otherwise. In the real-world industrial recommender systems, each user $u$ is associated with a user profile $x_u$ consisting of sparse features (\eg user id and gender) and numerical features (\eg user age), while each item $i$ is also associated with a item profile $x_i$ consisting of sparse features (\eg item id and brand) and numerical features (\eg price). 

In order to effectively incorporate auxiliary information of items (\ie item attributes and external knowledge) into recommendation, we frame our recommendation task over knowledge graph, which can be defined as follows:

\begin{myDef}
\textbf{Knowledge Graph}. A KG is defined as a directed graph $\mathcal{G} = \{\mathcal{E}, \mathcal{R}, \mathcal{T} \}$ with an entity set $\mathcal{E}$ and a relation set $\mathcal{R}$. Each triplet $(h, r, t ) \in \mathcal{T}$ denotes a fact that there is a relationship $r$ from head entity $h$ to tail entity $t$,where $h, t \in \mathcal{E}$ and $r \in \mathcal{R}$.
\end{myDef}

For example, $\langle Blouse, Category, Shirt \rangle$ states the fact that 
\emph{Blouse} belongs to the \emph{Shirt} \emph{Category}. To bridge knowledge graph with recommender system, we adopt a item-entity alignments function $\phi: \mathcal{I}  \rightarrow \mathcal{E}$ to align items with entities in KG.

Many efforts, especially GNN based methods, have been made to leveraging KG for better recommendation. While, most of these works overlook the mutual effect between target user and item when exploiting structural information derived from KG. To effectively distill structural knowledge through KG based GNN for item recommendation, we particularly investigate into the external knowledge connecting user behaviors and target item in KG, which can reveal semantic context for user-item interactions. Formally, we define such context information as follows:
\begin{myDef}
\textbf{Adaptive Target-Behavior Relational Graph}. Given a target user-item pair $\langle u, i \rangle$ and corresponding user behaviors $\mathcal{B}_{ui}$, an adaptive target-behavior relational graph \wrt $\langle u, i \rangle$ is defined as a sub-graph extracted from the original KG, connecting user behavior $\mathcal{B}_{ui}$ and target item $i$.
\end{myDef}

Given the above preliminaries, we now formulate the recommendation task to be addressed in this paper:
\begin{myDef}
\textbf{Task Description}. Given a knowledge graph $\mathcal{G}$ with historical interaction records $\mathcal{H}$, for each user-item pair $\langle u, i \rangle$, we aim to predict probability $\hat{y}_{ui}$ that user $u$ would click item $i$.
\end{myDef}

\section{The Proposed Framework}

\begin{table}
  \caption{Notations.}
  \label{table:notations}
  \begin{tabular}{ll}
    \toprule
    Notations & Description\\
    \midrule
    $\mathcal{U}$, $\mathcal{I}$ & \makecell[{}{p{5.8cm}}]{the set of users and items, respectively}\\
    $\mathcal{H}$ & \makecell[{}{p{5.8cm}}]{the set of historical interaction  records}\\
    $y_{ui}, \hat{y}_{ui}$ & \makecell[{}{p{5.8cm}}]{the label and the predicted probability}\\
    $x_u$, $x_i$, $\mathcal{B}_{ui}$ & \makecell[{}{p{5.8cm}}]{user profile, item profile and user behaviors}\\
    $i_b$ & \makecell[{}{p{5.8cm}}]{the specific item in  user behavior }\\
    $\mathcal{G}, \mathcal{G}_{ui}$ & \makecell[{}{p{5.8cm}}]{the knowledge graph and adaptive target-behavior relational graph, respectively}\\
    $\mathcal{E}, \mathcal{R}, \mathcal{T}$ & \makecell[{}{p{5.8cm}}]{the set of entity, relation and triples in knowledge graph, respectively}\\
    $\langle h, r, t\rangle$ & \makecell[{}{p{5.8cm}}]{the specific triple in knowledge graph}\\
    
    $\mathbf{x}_u, \mathbf{x}_i, \mathbf{x}_e, \mathbf{x}_r$  & \makecell[{}{p{5.8cm}}]{the embedding of user $u$, item $i$, entity $e$ and relation $r$, respectively } \\
    
    $\mathcal{N}_{h}^{(l)}$ & \makecell[{}{p{5.8cm}}]{the neighbors set in $l$-th relation-aware extractor layer for entity $h$}\\
    
    $\widetilde{\mathbf{x}}_i, \widetilde{\mathbf{x}}_{ib}$& \makecell[{}{p{5.8cm}}]{the relational representation for target item $i$ and each item $i_b \in \mathcal{B}_{ui}$, respectively} \\
    $\widetilde{\mathbf{x}}_u$& \makecell[{}{p{5.8cm}}]{the final representation of user $u$} \\
    \bottomrule
  \end{tabular}
\end{table}

    
\begin{figure*}
    \centering
    \includegraphics[scale=0.44]{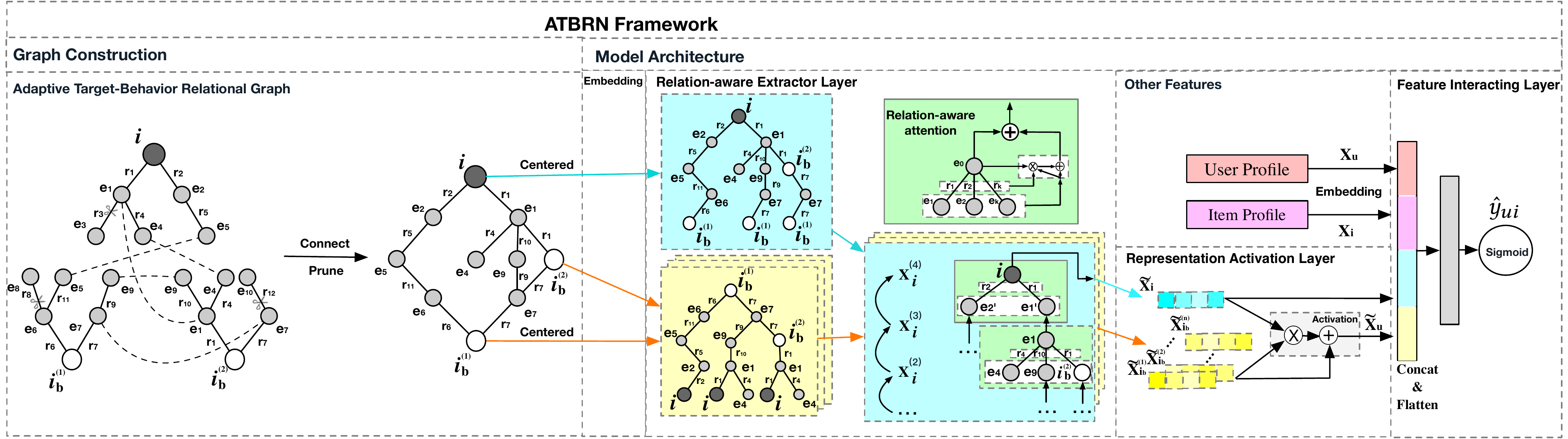}
    \caption{Overview of the proposed {\model} framework. Overall, {\model} consists of two parts, graph construction and model architecture.}
    \label{fig:framework}
\end{figure*}
In this section, we introduce our proposed framework {\model}, which aims to take full advantage of knowledge graph for recommendation. The framework is shown in Fig.~\ref{fig:framework}, which is composed of two modules: (1) To effectively extract structural relational knowledge for recommendation, we propose to construct the adaptive target-behavior relational graph for the given target user-item pair over knowledge graph, where the \textit{graph connect} and \textit{graph prune} techniques help mine high-order connective structure in an automatic manner;
(2) To jointly distill such a relational graph and rich relations among user behaviors in an end-to-end framework,  we elaborate on the model design of {\model}, which propagates user preference on the sub-graph with relation-aware extractor layer and representation activation layer. The key notations we will use throughout the article are summarized in Table \ref{table:notations}.


\subsection{Graph Construction}
A major novelty of our work is to effectively explore adaptive target-behavior relational graph for improving the modeling of the interaction. In this part, we introduce the strategy to construct the adaptive target-behavior relational graph with the proposed \textit{graph connect} and \textit{graph prune} techniques. To model the relationship between the given target user-item over KG, previous works either extract different paths through random walk~\cite{Wang:KPRN}, or directly leverage the neighbors of target item over the original KG~\citeN{Wang:KGAT,Wang:KGCN}.
Unfortunately, the first strategy neglects the structural relational information of the KG, while the second ignores the mutual effect between user behaviors and target item. Hence, we argue that above two strategies only achieve the suboptimal performance for recommendation.


Intuitively, the reasons driving a user to click a target item maybe implied by his/her historical behaviors, which is expected to guide our model to adequately aggregate useful information over external KG in an automatic manner. To distill the structural relational information over the KG in a more effective way, we propose to construct the adaptive relational graph w.r.t. user behaviors and target item. The procedure of the graph construction is clearly presented in the Algorithm \ref{alg:graph_construction} and left part of Fig. \ref{fig:framework}. Specifically, given a target user-item pair $\langle u, i \rangle$, we firstly exhaustively search the multi-layer entity neighbors for user behaviors $\mathcal{B}_{ui}$ and target item $i$ over the KG, and restore the paths connecting the entity and item into $\mathcal{G}_{ui}$ (lines 1-6). Through this, we connect the user behaviors and target item by multiple overlapped entities. Afterwards, for the entities in $\mathcal{G}_{ui}$, we prune the entities which do not connect different items. (lines 7-16). Finally, we get the relational graph $\mathcal{G}_{ui}$ for user $u$ and target item $i$, which describes the structural relations for $\langle u, i \rangle$ over the KG.

\begin{algorithm}
\caption{Graph construction} 
\label{alg:graph_construction} 
\begin{algorithmic}[1] 
\Require 
Target item $i$; User behavior $\mathcal{B}_{ui}$; Knowledge graph $\mathcal{G}$;
\Ensure 
$\mathcal{G}_{ui}$: Adaptive target-behavior relational graph for $\langle u, i \rangle$;
\For{item $v$ $\in$ [$i$, $\mathcal{B}_{ui}$]}:
    \For{entity $e \in \phi(v)$}:
        \State Construct path $p$ = ($e$, $r_k$, $e_k$, ..., $e_1$, $r_1$, $v$);
        \State $\mathcal{G}_{ui}$[$entity$] $\leftarrow$ $\mathcal{G}_{ui}[entity] \cup p$;\algorithmiccomment{Graph connect.}
    \EndFor
\EndFor
\For{entity $e \in \mathcal{G}_{ui}$}:
    \State New item hash set $s$;
    \For{path $p \in \mathcal{G}_{ui}[e]$}:
        \State Collect item $v$ on the path;
        \State $s \leftarrow s \cup v$;
    \EndFor
    \If{$s.size = 1$}:
        \State Prune $e$ in $\mathcal{G}_{ui}$; \algorithmiccomment{Graph prune.}
    \EndIf
\EndFor
\end{algorithmic}
\end{algorithm}

\subsection{Model Architecture}




After obtaining the adaptive target-behavior relational graph derived from the KG, we continue to study how to produce predictive embeddings for target user-item pairs through propagating user preference over such a sub-graph. 
As shown in the right part of Fig. \ref{fig:framework}, the model architecture of our proposed {\model} is composed of four layers: 
1) Embedding layer, which transforms high-dimensional sparse features into low-dimensional dense representations; 
2) Relation-aware extractor layer, which produces knowledge aware embeddings for user behaviors and target item by aggregating structural relational information over adaptive target-behavior relational graph;
3) Representation activation layer, which activates the relative relational representations of user behaviors w.r.t. that of target item. 
4) Feature interaction layer, which combines the user and item profile with the activated relational representation of user behaviors and target item for interaction.

\subsubsection{Embedding Layer}
As mentioned above, users and items in real-world recommendation scenario are both associated with abundant profile information in the form of sparse and dense features. Hence, we set up a embedding layer to parameterize users and items as vector representations, while preserving the above profile information. Formally, giving a user $u$, we have corresponding raw feature space $x_u$, comprised of sparse feature space $x^s_u$ and dense feature space $x^d_{u}$. For sparse features, following~\citeN{Paul:YoutubeNet,Zhou:DIN,zhou2019deep,Feng:DSIN}, we embed each feature value into $d$ dimensional dense vector, while dense feature can be standardized or batch normalization to ensure normal distribution. Subsequently, each user $u$ can be represented as $\mathbf{x}_u \in \mathbb{R}^{|x^s_u| \times d + |x^d_u|}$, where $|x^s_u|$ and $|x^d_u|$ denotes the size of sparse and dense feature space of user $u$, respectively. Similarly, we represent each item $i$ as $\mathbf{x}_i \in \mathbb{R}^{|x^s_i| \times d + |x^d_i|}$. Moreover, each entity $e$ and relation $r$ in the adaptive target-behavior relational graph can also be embeded as $\mathbf{x}_e \in \mathbb{R}^d$ and $\mathbf{x}_r \in \mathbb{R}^d$~\footnote{Entities and relations in KG only have the sparse id features}.


\subsubsection{Relation-aware Extractor Layer}
This layer is designed to effectively and comprehensively distill the structural relational information from the extracted sub-graph.
Previous works~\citeN{Veli:GAT, Kipf:GCN} neglect relational edges during aggregation, which play essential roles in  real-world settings. In our scenario, a user $u$ may \textit{click} or \textit{buy} the same item $i$, while the relations \textit{click} and \textit{buy} obviously indicate the different preference of user $u$ towards item $i$. Therefore, we elaborately build the relation-aware extractor layer to adequately exploit rich structural relational information in KG in the consideration of various relation between entities.

Based on the above discussions and inspired by the study \cite{Gong:EGNN}, we stack the relation-aware extractor layer by layer in order to recursively propagates the embeddings from an entity's neighbors to refine the entity's embedding in KG. Specifically, for each item (\ie item $i_b \in \mathcal{B}_{ui}$ or target item $i$), we will regard it as the center node and aggregate information over the extracted sub-graph $\mathcal{G}_{ui}$ through relation-aware aggregation. Given an entity~\footnote{For convenience, we omit the subscript $_{ui}$ in this part.} $h$ in extracted relational sub-graph $\mathcal{G}_{ui}$,  let $\mathcal{N}^{(l)}_h = \{(r,t)|(h,r,t) \in \mathcal{G}_{ui}\}$ to denote the neighbors set in $l$-th layer and $\mathbf{x}_h^{(l)}$ to denote the  representation of entity $h$ in the $l$-th layer. We implement the $l$-th relation-aware aggregation layer as follows,
\begin{equation}\label{EQ1}
    \begin{split}
    \alpha^{(l)}(h, r, t) &= \frac{\exp(\mathbf{x}_r\mathbf{W}_{\alpha}f(\mathbf{x}^{(l)}_h\oplus\mathbf{x}^{(l)}_t))}{\sum_{(r',t') \in \mathcal{N}^{(l)}_h} \exp(\mathbf{x}_{r'}\mathbf{W}_{\alpha}f(\mathbf{x}^{(l)}_h\oplus\mathbf{x}^{(l)}_{t'}))}, \\
    \textbf{x}_h^{(l+1)} &= \mathbf{x}_h^{(l)} \oplus \sum_{(r,t) \in \mathcal{N}^{(l)}_h} \alpha^{(l)}(h, r, t)\mathbf{x}^{(l)}_t.\\
    \end{split}
\end{equation}
Here, $f(\mathbf{x})$
and $\mathbf{W}_{\alpha}$ denote the single layer perceptron and attentive matrix in $l$-th layer, respectively. And $\oplus$ denotes the concatenation operation. Relation-aware extractor layer is stacked layer by layer to propagate user preference over KG. Subsequently, each entity $e$ in sub-graph $\mathcal{G}_{ui}$ can be denoted as $\mathbf{x}^{(L)}_e$ after $L$ relation-aware extractor layer.

Given a target user-item pair $\langle u, i \rangle$ and the corresponding adaptive target-behavior relational graph $\mathcal{G}_{ui}$, the knowledge aware representation of target item $i$ can be denoted as $\widetilde{\mathbf{x}}_{i} = \mathbf{x}^{(L)}_{i}$, where $\phi(i) = e$. Similarly, we also obtain relational representation set $\{ \widetilde{\mathbf{x}}_{i_b} = \mathbf{x}^{(L)}_{i_b} | \phi(i_b) = e\}_{i_b \in \mathcal{B}_{ui}}$ for user behaviors.

\subsubsection{Representation Activation Layer}\label{section323}
Intuitively, user behaviors contribute differently to the final prediction. 
For example, the behavior $shirt\ A$ is more informative than  $shoe\ B$ when the target item is $shirt\ C$. For this purpose, we set a representation activation layer to place different importance on relational representation of user behaviors $\{ \widetilde{\mathbf{x}}_{i_b}\}_{i_b \in \mathcal{B}_{ui}}$. Specifically, we apply the vanilla attention mechanism~\cite{bahdanau2014neural} to activate representations of user behaviors that are more related to target item, calculated as follows,
\begin{equation}\label{EQ2}
    \begin{split}
    \beta(u, i, i_b) &= \frac{\exp(\widetilde{\mathbf{x}}_{i_b}\mathbf{W}_{\beta}\widetilde{\mathbf{x}}_{i}))}{\sum_{i'_b \in \mathcal{B}_{ui}} \exp(\widetilde{\mathbf{x}}_{i'_b}\mathbf{W}_{\beta}\widetilde{\mathbf{x}}_{i})}, \\
    \widetilde{\mathbf{x}}_u &= \sum_{i'_b \in \mathcal{B}_{ui}} \beta(u, i, i_b)\widetilde{\mathbf{x}}_{i'_b},\\
    \end{split}
\end{equation}
where $\mathbf{W}_{\beta}$ is the attentive matrix in representation activation layer.


\subsubsection{Feature Interaction Layer}
Until now, given a target user-item pair $\langle u, i\rangle$, we have the profile embeddings for user $u$ and item $i$, and knowledge aware embedding from adaptive target-behavior relational graph for user behaviors and target item. We combine the four embedding vectors into a unified representation and employ Multiple Layer Perceptron (MLP) for better feature interaction~\citeN{Zhou:DIN,Zhou:ATRank,Feng:DSIN,Pi:MIMN,zhou2019deep}. 
\begin{equation}\label{EQ3}
    \begin{split}
    \hat{y}_{ui} = \sigma(f(f(f(\mathbf{x}_u \oplus \mathbf{x}_i \oplus \widetilde{\mathbf{x}}_{u} \oplus \widetilde{{\mathbf{x}}_i})))),
    \end{split}
\end{equation}
where $\sigma(\cdot)$ is the logistic function and $\hat{y}_{ui}$ represents the prediction probability of the user $u$ to click on the target item $i$.


\subsubsection{Loss Function}
We reduce the task to a binary classification problem and use binary cross-entropy loss function defined as follows:
\begin{equation} 
    \begin{split}
    \mathcal{L} = - \frac{1}{N} \sum_{(u, i) \in \mathbb{D}} (y_{ui}\,\log \hat{y}_{ui} + (1 - y_{ui})\,\log(1 - \hat{y}_{ui}))
    \end{split}
    \label{eq:loss function}
\end{equation}
where $\mathbb{D}$ is the training dataset and $y_{ui} \in \{0, 1\}$ represents whether the user $u$ clicked on the target item $i$.


\section{Experiments}
In this section, we perform a series of experiments on two real-world datasets, with the aims of answering the following research questions:
\begin{itemize}
    \item \textbf{RQ1}: How does our proposed model {\model}  perform compared with state-of-the-art methods on the recommendation task?
    \item \textbf{RQ2}: How do different experimental settings (i.e., depth of graph, aggregator selection, etc.) influence the performance of {\model}?
    \item \textbf{RQ3}: How does {\model} provide effective recommendation intuitively?
\end{itemize}

\begin{table}
\caption{Statistics of datasets}
\label{table:dataset}
\begin{tabular}{c|c|c|c}
\hline
{} & {Description} &  {Taobao} & {Yelp}\\
\hline
\multirow{3}{*}{} & {\#Users} & {2.2 $\times 10^8$} & {4.5 $\times 10^4$} \\
{User-Item} & {\#Items} & {1.1 $\times 10^8$} & {4.5 $\times 10^4$} \\
{Interaction} & {\#Interactions} & {7.2  $\times 10^9$} & {1.0 $\times 10^6$} \\
\hline 
\multirow{3}{*}{Knowledge}& {\#Entities} & {1.4 $\times 10^7$} & {8.3 $\times 10^4$} \\
{}& {\#Relations} & {34} & {35} \\
{Graph}& {\#Triplets} & {3.8 $\times 10^{10}$} & {1.6 $\times 10^6$} \\
{}& {\#Max neighbor depth} & {3} & {1} \\
\hline
\end{tabular}
\end{table}

\subsection{Experimental Setup}
\subsubsection{Datasets}
We conduct extensive experiments on two real-world datasets: industrial dataset from Taobao and benchmark dataset from Yelp. 
\begin{itemize}
    \item \textbf{Taobao}~\footnote{www.taobao.com.} dataset consists of click logs from 2019/08/22 to 2019/08/29, where the first one week's samples are used for training and samples of the last day are for testing. Moreover, Taobao dataset also contains user profile (\eg id and age), item profile (\eg id and category) and up to 10 real-time user behaviors~\footnote{Real-time user behaviors means user behaviors before this action occurs.}.
    \item \textbf{Yelp}~\footnote{www.yelp.com/dataset.} dataset records interactions between users and local business and contains user profile (\eg id, review count and fans), item profile (\eg id, city and stars) and up to 10 real-time user behaviors. For each observed interaction, we randomly sample 5 items that the target user did not engage with before as negative instances. For each user, we hold the latest 30 instances as the test set and utilizes the remaining data for training.
\end{itemize}

Besides user behaviors, following~\citeN{luo2020alicoco,Shen:EntityLinking,Zhao:KB4Rec}, we construct item knowledge for Taobao (\eg category, parent and style). Also, for Yelp dataset, KG is organized as the local business information (e.g., location and category). 
The detailed descriptions of the two datasets are shown in Table~\ref{table:dataset}. Note that the volume of Taobao dataset is much larger than yelp, which brings more challenges.
\subsubsection{Baselines} 
We compare our {\model} with three kinds of representative methods: feature based methods (\ie YoutubeNet and DeepFM) mainly utilizing raw features derived from user and item profile, behavior based methods (\ie DIN, DIEN and DSIN) capturing user's historical behaviors and knowledge graph (KG) based methods (\ie RippleNet, KGAT and KPRN) benefiting from knowledge graphs in recommendation. The comparison methods are given below in detail:
\begin{itemize}
    \item \textbf{YoutubeNet}~\cite{Paul:YoutubeNet} is a standard user behavior based method in the industrial recommender system. 
    \item \textbf{DeepFM}~\cite{Guo:DeepFM} combines factorization machine and deep neural network for recommendation.
    \item \textbf{DIN}~\cite{Zhou:DIN} locates related user behaviors \wrt target itemby using attention mechanism.
    
    \item \textbf{DIEN}~\cite{zhou2019deep} models users temporary interests and the interest evolving process via GRU with attention update gate.
    \item \textbf{DSIN}~\cite{Feng:DSIN} models user's session interests and the evolving process with self-attention mechanism and Bi-LSTM.
    \item \textbf{RippleNet}~\cite{wang2018ripplenet} propagates user's potential preferences over the set of knowledge entities.
    
    \item \textbf{KPRN}~\cite{Wang:KPRN} is a typical path based recommendation method, which extracts qualified path to between a user with an item.
    
    \item \textbf{KGAT}~\cite{Wang:KGAT} is a state-of-the-art KG-based recommendation methods, which employs GNN on KG to generate representations of users and items, respectively.
    
\end{itemize}

\subsubsection{Evaluation Metrics} We adopt area under ROC curve (\textbf{AUC}) to evaluate the performance of all methods.  
Larger AUC indicates better performance. Besides, we also present the relative improvement (\textbf{RI}) \wrt AUC of our model achieves over the compared models, which can be formulated as:
\begin{equation}
    \text{RI} = \frac{|\text{AUC}(model) - \text{AUC}(base)|}{\text{AUC}(base)} * 100\%,
\end{equation}
where $|.|$ is the absolute value, $model$ refers to our proposed framework {\model} and $base$ refers to the baseline. Note that \textbf{0.001} improvement \wrt AUC is remarkable in industrial scenario (\ie Taobao dataset).



\subsubsection{Implementation}
We implement all models in Tensorflow 1.4. Moreover, for fair comparison, pre-training, batch normalization and regularization are not adopted in our experiments. For RippleNet, we set the max depth of ripple as 3. For KGAT, the max neighbour depth of target user and item is set to 4 and 3, respectively. For KPRN, the max number of extracted paths over the knowledge graph are set to 50. For {\model}, the max neighbor depth of the item is set to 3. For all models, We employ random uniform to initialize model parameters and adopt Adagrad as optimizer using a learning rate of 0.001. Moreover, embedding size of each feature is set to 4 and the architecture of MLP is set to [512, 256, 128].
We run each model three times and reported the mean of results.

\subsubsection{Significance Test}
For Experimental results in Tables 4, 5, 6 and 7, we use ``*'' to indicate that {\model} is significantly different from the runner-up method based on paired t-tests at the significance level of 0.01.

\begin{table}
\caption{Overall performance comparison \wrt AUC (bold: best; underline: runner-up). }
\label{table:model_performance}
\begin{tabular}{c|c|c|c|c}
\hline
\multirowcell{2}{Model} & \multicolumn{2}{{c|}}{Taobao} & \multicolumn{2}{{c}}{Yelp$^\dag$} \\
\cline{2-5}
{} & {AUC} & {RI} & {AUC} & {RI} \\
\hline 
{YoubtubeNet} & {0.6017}  & {+2.72\%} & {0.7109} & {+26.00\%} \\
{DeepFM} & {0.6037}  & {+2.38\%} & {0.7334} & {+22.14\%} \\
\hline
{DIN} & {0.6058} & {+2.03\%} & {0.7520} & {+19.12\%} \\
{DIEN} & {0.6061}  & {+1.97\%} & {0.7581} & {+18.16\%} \\
{DSIN} & {0.6073}  & {+1.77\%} & {0.7774} & {+15.23\%} \\
\hline
{RippleNet} & {0.5975} & {+3.44\%} & {0.7324} &{+22.31\%} \\
{KGAT} & {0.6062}  & {+1.96\%} & {0.7876} & {+13.73\%} \\
{KPRN} & {\underline{0.6096}}  & {+1.39\%} & {\underline{0.8260}} & {+8.45\%} \\
\hline
{{\model}} & {\textbf{0.6181}$^\ast$}  & {-} & {\textbf{0.8958}$^\ast$} & {-} \\
\hline
\end{tabular}
\begin{tablenotes}
\item[$\dag$] $^\dag$ Note that the relative improvement on the public Yelp dataset is much higher than the industrial Taobao dataset, since the negative samples of the public Yelp dataset are generated by random sampling and thus easier to distinguish.
\end{tablenotes}
\end{table}

\subsection{Performance Comparison (RQ1)}
We report the AUC comparison results of {\model} and baselines on two datasets in Table~\ref{table:model_performance}. The major findings from the experimental results are summarized as follows:

\begin{itemize}
    \item Feature based methods (\ie YoutubeNet and DeepFM) achieve relatively pool performance on two datasets. It indicates that handcrafted feature engineering is insufficient to capture the complex relations between users and items, further limiting performance. Moreover, DeepFM consistently outperforms YoutubeNet across all cases, since it employs FM part for better feature interaction.
    
    \item Compared to feature based methods, the performance of behavior based methods (\ie DIN, DIEN and DSIN) verifies that incorporating historical behaviors is beneficial to infer user's preference. Among them, DSIN achieves the best performance on both datasets due to integration of user's session interests.
    
    \item Generally, KG based methods (\ie RippleNet, KGAT, KPRN) achieve better performance than behavior based methods in most cases, which indicates the effectiveness of knowledge graph for capturing underlying interaction between users and items. However, RippleNet underperforms other baselines on both datasets. One possible reason is that RippleNet ignores user's short-term interest implied in historical behaviors. Moreover, KPRN generally achieves remarkable improvements in most cases. It makes sense since reasonable and explainable target user-item paths extracted from KG are helpful to improve recommendation performance.
    
    \item {\model} consistently yields the best performance on both datasets. In particular, {\model} improves over the best baseline w.r.t. AUC by 1.39\%, and 8.45\% on Taobao and Yelp dataset, respectively. 
    By stacking multiple GNN layers, {\model} is capable of exploring rich structural and relational information over KG, while KPRN only models each extracted path independently. This verifies the importance of capturing both semantics and topological structures derived from KG for recommendation.
    Besides, compared with KGAT, which only represents target user and item separately by aggregating their own neighbors over the original KG, {\model} achieves better performance for the following two reasons: 1) {\model} considers the mutual effect between the given user behaviors and target item by constructing the adaptive relational sub-graph for them. Propagating on such a sub-graph can better capture the structural relations between user behaviors and target item and further explore potential reasons driving the user to click the target item; 2) {\model} integrates relations when aggregating the entities by the relation-aware attention mechanism, and creatively produces the relational representations over the extracted sub-graph for each user behavior and target item. 
    

\end{itemize}

\subsection{Study of {\model} (RQ2)}


\begin{table}
\caption{Effect of the representation activation layer and relation-aware mechanism.}
\label{table:ablation_study_1}
\begin{tabular}{c|c|c|c|c}
\hline
\multirowcell{2}{Model} & \multicolumn{2}{{c|}}{Taobao} & \multicolumn{2}{{c}}{Yelp} \\
\cline{2-5}
{} & {AUC} & {RI} & {AUC} & {RI} \\
\hline
{{\model}$_{w/o\ RAM}$}  & {0.6157} & {+0.38\%}& {0.8858}& {+1.12\%}\\
{{\model}$_{w/o\ RAL}$} & {0.6125} & {+0.91\%}& {0.8940}& {+0.20\%}\\
\hline
{\model} & \textbf{0.6181$^\ast$} & {-} & \textbf{0.8958$^\ast$} & {-} \\
\hline
\end{tabular}
\end{table}

\begin{table}
\caption{Effect of the depth of neighbor.}
\label{table:ablation_study_2}
\begin{tabular}{c|c|c|c|c}
\hline
\multirowcell{2}{Model$^\dag$} & \multicolumn{2}{{c|}}{Taobao} & \multicolumn{2}{{c}}{Yelp} \\
\cline{2-5}
{} & {AUC} & {RI} & {AUC} & {RI} \\
\hline
{\model}$_{1/0}$  & 0.6054 &{+2.09\%} & 0.7523 &{+19.07\%} \\
{\model}$_{3/1}$  & 0.6143 &{+0.61\%} &\textbf{0.8958$^\ast$} &{-} \\
{\model}$_{5/2}$  & \textbf{0.6181$^\ast$} &{-} & - &{-}\\
{\model}$_{7/3}$  & 0.6163 &{+0.29\%} & - & {-} \\
\hline
\end{tabular}
\begin{tablenotes}
\item[$\dag$] $\dag${\model}$_{m/n}$ means {\model} explores $m$ layers of neighbors over the extracted sub-graph, which corresponds to $n$ layers of neighbors in original KG.
\end{tablenotes}
\end{table}

\begin{table}
\caption{Effect of different aggregators.}
\label{table:ablation_study_3}
\begin{tabular}{c|c|c|c|c}
\hline
\multirowcell{2}{Model} & \multicolumn{2}{{c|}}{Taobao} & \multicolumn{2}{{c}}{Yelp}  \\
\cline{2-5}
{} & {AUC} & {RI} & {AUC} & {RI} \\
\hline
{\model}$_{concat}$ & 0.6131 & {+0.64\%} & 0.8901  & {+0.65\%}\\
{\model}$_{sum}$& 0.6133 & {+0.78\%} & 0.8906  & {+0.58\%}\\
{\model}$_{sa}$ & 0.6145  & {+0.58\%}& 0.8928  & {+0.33\%}\\
{\model}$_{nl}$ & 0.6159  & {+0.35\%}& 0.8946 & {+0.13\%} \\
\hline
{\model} & \textbf{0.6181$^\ast$}  & {-}& \textbf{0.8958$^\ast$}  & {-}\\
\hline
\end{tabular}
\end{table}

In this section, we perform a series of experiments to better understand the traits of {\model}, including well-designed components (\eg relation-aware mechanism and representation activation layer) and key parameter settings (\ie neighrbor depth and aggregator).

\subsubsection{Effect of Relation-aware Mechanism and Representation Activation Layer} {\model} provides a principled way to characterize various relations in KG and user behaviors to enhance recommendation performance. To examine the effectiveness of relation-aware mechanism and representation activation layer, we prepare three variants of {\model}:
\begin{itemize}
    \item \textbf{\model}$_{w/o\ RAM}$: The variant of {\model}, which removes the relation-aware mechanism (Eq.~\ref{EQ1}).
    \item \textbf{\model}$_{w/o\ RAL}$: The variant of {\model}, which removes the representation activation layer (Eq.~\ref{EQ2}).
\end{itemize}

The AUC comparison results of {\model} with its variants are show in Table \ref{table:ablation_study_1}. We have the following two observations: 
\begin{itemize}
    \item It is clear that the performance of {\model} degrades without the relation-aware mechanism on both datasets (\ie {\model} $>$ {\model}$_{w/o\ RAM}$). It demonstrates that different relations in KG should be distinguished, as disregarding such information leads to the worse performance.
    \item {\model} without the representation activation layer performs worse consistently (\ie {\model} $>$ {\model}$_{w/o\ RAL}$). It indicates that capturing the semantic relations among user behaviors over the KG can better understand user underlying preference, which is beneficial for the final prediction.
\end{itemize}

\begin{figure}
    \centering
    \includegraphics[scale=0.25]{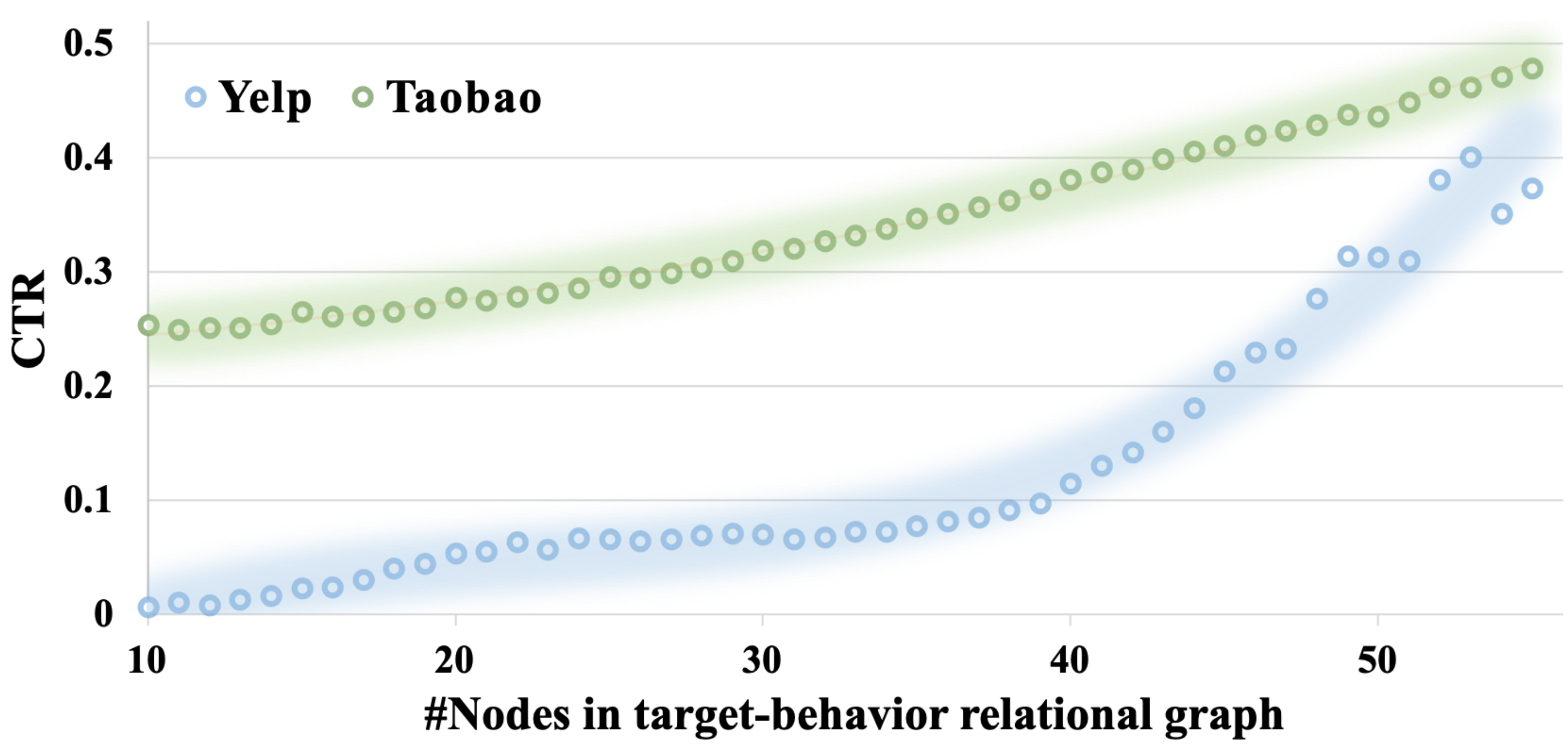}
    \caption{Impact of the number of nodes in adaptive target-behavior relational graph \wrt CTR.}
    \label{fig:subgraph}
\end{figure}

\begin{figure*}
    \centering
    \includegraphics[scale=0.38]{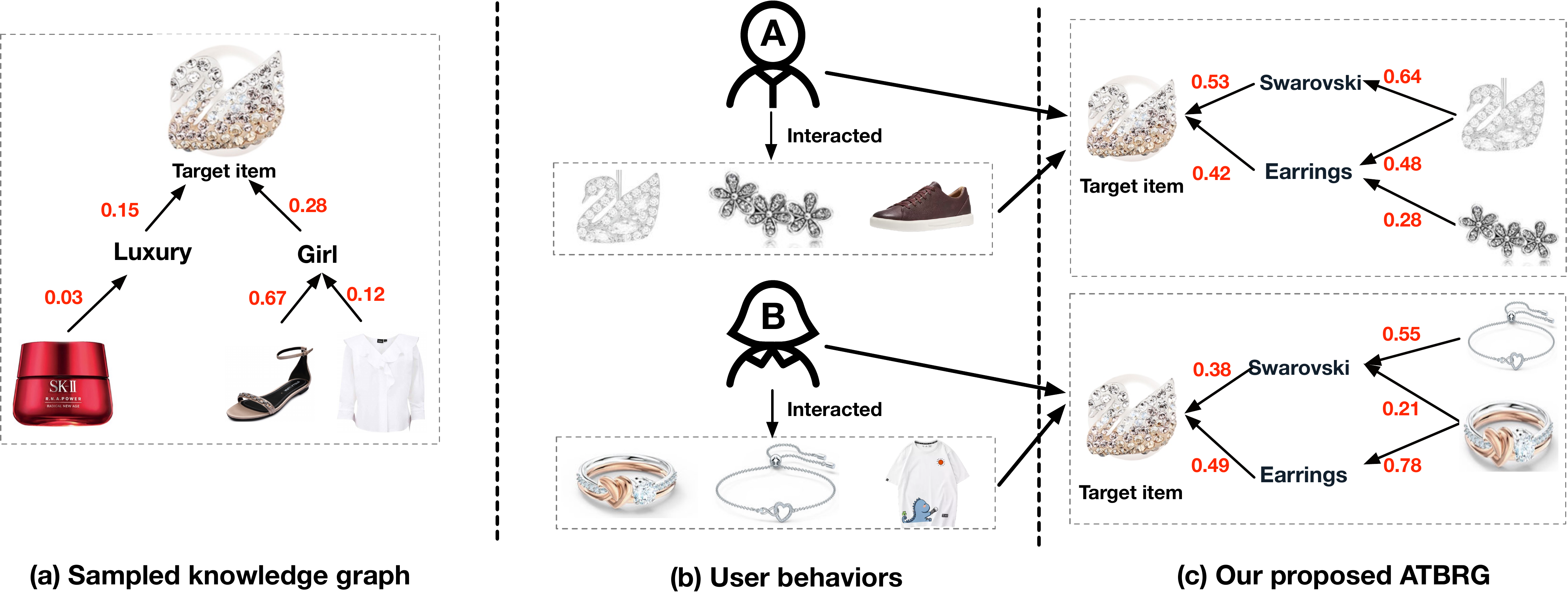}
    \caption{An illustrative example of how our proposed {\model} works more effective than other knowledge graph based methods. (a) indicates the sampled neighbors of the target item over the knowledge graph. (b) \& (c) introduce the specific user behaviors of users and the corresponding extracted sub-graph. The red number above the edge implies the calculated weights.}
    \label{fig:case_study}
\end{figure*}


\subsubsection{Effect of Neighbor Depth}
The proposed {\model} model is flexible to capture high-order structural information through recursively aggregating the embeddings from an entity's neighbors to refine the entity's embedding in KG. Here, we investigate how the neighbor depth over KG influences the model performance. Specifically, the neighbor depth of item is explored in the range of \{0, 1, 2, 3\} in Taobao dataset and \{0, 1\} for Yelp dataset. We summarize the results in Table~\ref{table:ablation_study_2} and have the following two observations: 
\begin{itemize}
    \item Overall, the model performance increases gradually when neighbor depth varies from 0 to 2 on both datasets. It demonstrates that deepening the neighbor layer helps capture the long-term structural relations of user behaviors and target item to some extent.
    \item The model performance of {\model} degrades when the neighbor depth increases from 2 to 3. One possible reason is that long-term relations may include much more ineffective connectives (\ie $Shirt$ - $Women\ Clothing$ - $Clothing$ - $Men\ Clothing$ -$Shoe$). Such relations in the graph introduce some noise and further harm the model performance.
\end{itemize}


\subsubsection{Effect of Aggregator}
In our model, we enrich the information of items by recursively capturing their neighbor information in KG. In order to explore the effect of different neighbor aggregator, we design different variants of {\model}, listed as follows:
\begin{itemize}
    \item \textbf{\model}$_{concat}$: It applies concatenation operation~\cite{Wang:KGCN}.
    \item \textbf{\model}$_{sum}$: It applies sum pooling.
    \item \textbf{\model}$_{sa}$: It applies self-attention mechanism~\cite{Veli:GAT}.
    \item \textbf{\model}$_{nl}$: It applies nonlinear transformation~\cite{Wang:KGCN}.
\end{itemize}

We present the AUC comparison of {\model} and its variants in Table \ref{table:ablation_study_3}. From the result, we have the following findings: 
\begin{itemize}
    \item Obviously, {\model} with simple aggregators (\ie {\model}$_{concat}$ and {\model}$_{sum}$) performs worst on both datasets, since they ignore the different contributions of neighbors.
    \item Generally, {\model} with complex aggregators (\ie {\model}$_{sa}$ and {\model}$_{nl}$) achieves better performance on both datasets. The reason is that  {\model}$_{sa}$ employs self-attention mechanism to place different importance on neighbors while {\model}$_{nl}$ leverages nonlinear trasformation to characterize complex interaction.
    \item {\model} consistently yields the best performance on both datasets. It illustrates that, our proposed relation-aware aggregator not only includes the nonlinear transformation in the weight calculation, but also considers the influence of relations during aggregation.
\end{itemize}


\subsection{Case Study (RQ3)}

To better understand the merits of our proposed {\model} intuitively, we first make comprehensive instance-level analyses on the adaptive target-behavior relational graph. As shown in Fig. \ref{fig:subgraph}, we present the influence of the number of nodes over the relational graph on the click through rate (CTR). Here, CTR is calculated by averaging the real labels (1 for click and 0 otherwise). It can be clearly observed that CTR and the number of nodes are positively correlated on both datasets.
This demonstrates that, the richer the relations between user behavior and target item on the extracted sub-graph, the more likely the user is to click on target item.

Moreover, with the aims of answering how {\model} addresses the limitations (described in Section 1) existed in previous GNN based methods for knowledge-aware recommendation, we conduct one case study in large-scale industrial Taobao dataset. As shown in Fig. \ref{fig:case_study}, the main findings are summarized as follows:
\begin{itemize}
    \item In part (a), due to the limitation (\textbf{L2}), the original neighbors of the target item over the knowledge graph are randomly sampled beforehand. Hence, some relevant entities (\ie $Swarovski$ and $Earring$) connecting users are discarded, while other ineffective entities (\ie $Luxury$ and $Girl$) are reserved, which inevitably introduce noises. It demonstrates that previous methods are incapable of adaptively sampling neighbors for target user-item pairs, and further harm the recommendation performance.
    \item In part (b), we present the users' recent behaviors, where some behaviors (\ie $Necklace$ and $Earring$) are related to target item over the knowledge graph while others (\ie $Shoe$ and $Skirt$) are not. By the graph connect and prune techniques, we adaptively preserve the effective entities and relations over knowledge graph (\textbf{L2}). Subsequently, in part (c), we construct the specific adaptive target-behavior relational graph for the given target user-item pair, which provides strong evidences for inferring user preference. Propagating embeddings on on such a relational structure can take full advantage of the mutual effect of target user-item pair for recommendation (\textbf{L1}).
    
    \item In order to consider the rich relations among user behaviors over KG, we propose the relation-aware extractor layer to weigh various underlying preferences for recommendation.
    Compared with part (a), we find the weights are also adaptive for different users (\textbf{L3}). Specifically, $Swarovski$ is paid more attention to and scored higher than $Earring$ by the user A, while it is the opposite for user B. Therefore, the final relational representation can reflect the personalized preferences of different users towards the target item.
\end{itemize}

\subsection{Online A/B Testing}
\begin{figure}
    \centering
    \includegraphics[scale=0.55]{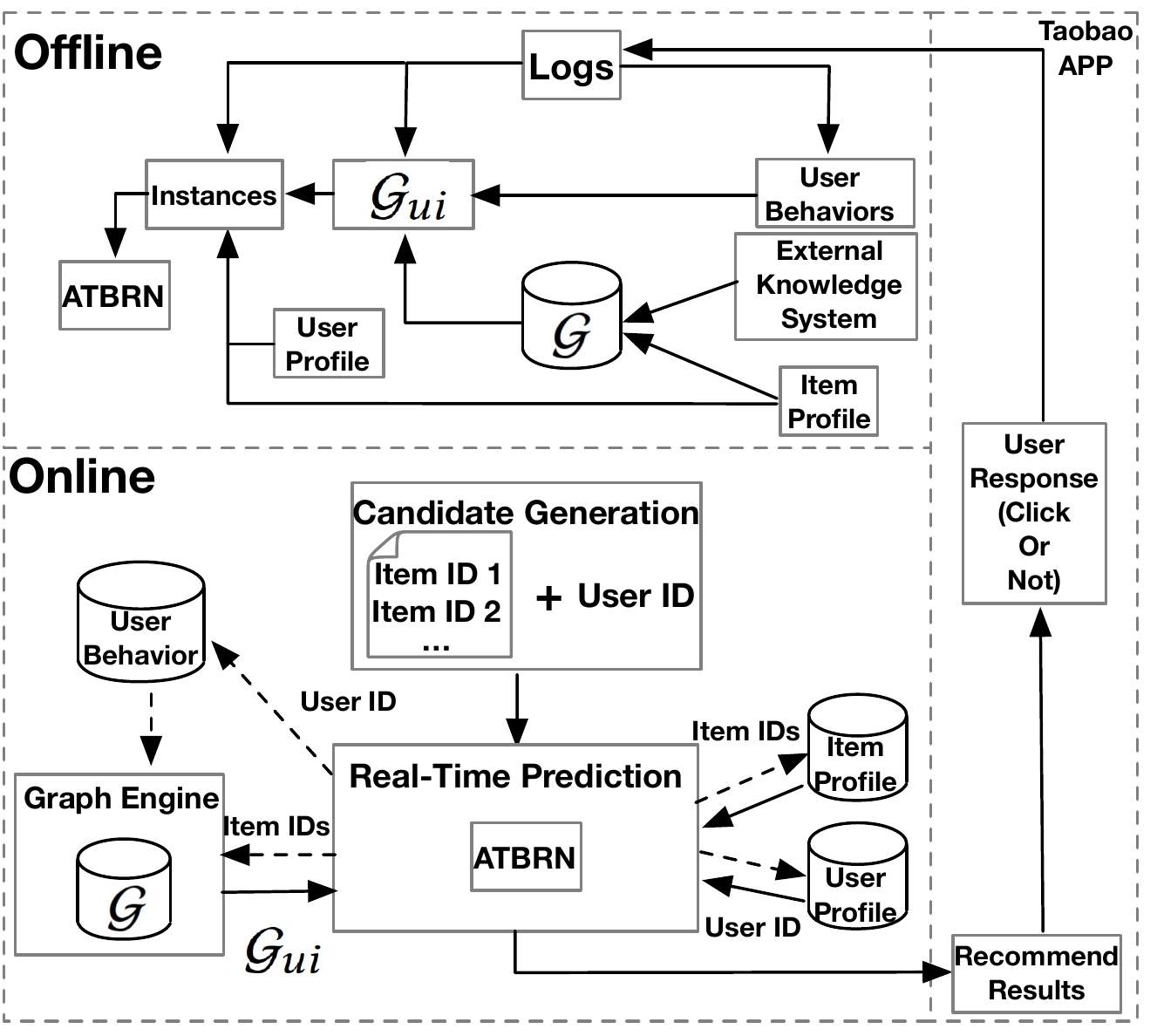}
    \caption{The deployment of {\model} in Taobao APP.}
    \label{fig:online_serving}
\end{figure}
To verify the effectiveness of our proposed framework {\model} in the real-world settings, {\model} has been deployed in the popular recommendation scenario of Taobao APP. As shown in Fig. \ref{fig:online_serving}, the deployment pipelines consist of three parts: 1) User response. Users give implicit feedback (click or not) to the recommended items provided by the recommender system; 2) Offline training. In this procedure, we 
integrate the knowledge graph $\mathcal{G}$, user behaviors and target item to construct the adaptive target-behavior relational graph $\mathcal{G}_{ui}$. Afterwards, $\mathcal{G}_{ui}$, together with user profile and item profile makes up the instances, and are fed into {\model} for training; 3) Online serving. When the user accesses Taobao APP, some candidates items are generated by the pipelines before real-time prediction (RTP) service. The necessary components of {\model} are achieved and organized in the same way as the offline training. At last, the candidates are ranked by the predicting scores of {\model}, and truncated for the final recommend results.

Compared with existed deployed baseline model DIN,  \textbf{6.8}\% lift on click count and \textbf{5.1}\% lift on CTR are observed for {\model}, with the cost of \textbf{8} milliseconds for online inference. The promotion of recommendation performance verifies the effectiveness of our proposed framework {\model}.

\section{Conclusion}
In this paper, we propose a novel framework {\model} for knowledge aware recommendation. To effectively characterize the structure relations over KG, we propose the graph connect and graph prune techniques to construct adaptive target-behavior relational graph. Furthermore, we elaborate on the model design of {\model}, equipped with relation-aware extractor layer and representation activation layer, which aims to take full advantage of structural connective knowledge for recommendation. Extensive experiments on both industrial and benchmark datasets demonstrate the effectiveness of our framework compared to several state-of-the-art methods. Moreover, {\model} has also achieved 5.1\% improvement on CTR metric in online experiments after successful deployment in one popular recommendation scenario of Taobao APP.
In the future, we will consider applying causal inference in KG to improve the interpretability of recommender system.


\bibliographystyle{ACM-Reference-Format}
\bibliography{references}

\end{document}